\documentclass[12pt]{iopart}
\usepackage{iopams}  
\usepackage{amsmath,amssymb,xcolor,graphicx,comment}
\usepackage[T1]{fontenc}
\begin{document}

\title{Classical signatures of quenched and thermal disorder in the dynamics of correlated spin systems}

\author{$^{1,2,3}$Harry Lane, $^{4}$Kipton Barros and $^{5}$Martin Mourigal}
\address{$^{1}$Department of Physics and Astronomy, School of Natural Sciences, The University of Manchester, Oxford Road, Manchester M13 9PL, UK.}
\address{$^{2}$The University of Manchester at Harwell, Diamond Light Source, Didcot Oxfordshire
OX11 0DE, UK.}
\address{$^{3}$School of Physics and Astronomy, University of St Andrews, St Andrews, United Kingdom KY16 9SS, UK.}
\address{$^4$Theoretical Division and CNLS, Los Alamos National Laboratory, Los Alamos, New Mexico 87545, USA}
\address{$^{5}$School of Physics, Georgia Institute of Technology, Atlanta, Georgia 30332, USA}
\ead{harry.lane@manchester.ac.uk}
\vspace{10pt}
\begin{indented}
\item[]January 2025
\end{indented}

\begin{abstract}
Neutron scattering is frequently used to look for evidence of features indicative of quantum-entangled phases of matter such as continua from fractionalisation or quantised excitations. However, the non-specificity of these features and difficulty of both fully quantum treatments and semiclassical models of disorder, make the diagnosis of such states problematic. Here, we demonstrate the feasibility of semiclassical treatments of disordered systems for supercells of $\sim 10,000$ spins. By examining a number of classically disordered models we show the presence of quantised excitations, broad continua and anomalous damping originating from quenched disorder or large classical degeneracies.      
\end{abstract}

\section{Introduction}
\label{Section:Introduction}

Disorder is ubiquitous in materials and takes many forms. It can be driven by classical thermal phenomena or quantum mechanical zero-point fluctuations. Within the realm of the impact that classical disorder has on magnetism, a distinction can be made between quenched disorder and dynamical disorder. 
Quenched disorder involves spatial heterogeneities, such as defects frozen during material synthesis, which remain static at the time scale of the magnetic phenomena. In contrast, dynamical disorder results from the intrinsic equilibrium dynamics of the magnetic moments (spins). This last phenomenon is significant for highly correlated paramagnets, commonly found in frustrated systems~\cite{Ramirez94:24}. In these systems, large thermal fluctuations persist for temperatures well below the characteristic interaction scale between spins. In extreme cases, such as classical spin liquids, zero modes associated with extensive ground-state degeneracies play a role in the dynamics, and metastable states can form around many local energy minima~\cite{Moessner98:80,Zhang19:122}.

Controllable quenched disorder in the form of chemical doping serves as a powerful tuning parameter in materials physics~\cite{Zunger21:121,Keimer17:13}. Even minute changes in carrier densities in metals can induce superconductivity~\cite{Lee06:78} or drive an instability to spatially modulated charge or spin density wave phases~\cite{Sebastian15:6,Ghiringhelli12:337,Peng18:17,Pratt11:106,Nandkishore12:86,Lee99:60,Slobodchikov22:106}. In magnetic insulators, it can alter the strength of single-ion anisotropies or stabilise magnetism at high temperatures~\cite{Zhang22:8,Meisenheimer23:14,Chen21:12}. Conversely, uncontrolled (often unknown) heterogeneities, such as crystal defects and stresses, have blighted the search for magnetic phases dominated by quantum fluctuations~\cite{Clark21:51}. Not only can disorder cause glass formation~\cite{Binder86:58,Syzranov22:13}, but it can also conspire to mimic the continuous spectra of fractionalised excitations usually associated with extreme quantum effects in neutron scattering experiments~\cite{Zhu17:119,Bai19:122,Conlon10:81,Lane2024:preprint,Zhang19:122}. Distinguishing between classical and quantum disorder effects through the static or dynamic spin structure factors, ${\mathcal{S}(\bf Q)}$ and ${\mathcal{S}(\bf Q,\omega)}$, measured in such experiments is a challenge because high-fidelity quantum mechanical calculations are currently limited to a small range of problems. The situation becomes even more complex in practice, because thermal and quantum disorder effects can induce ordering through mechanisms known as order-by-disorder.

The task of differentiating between the effects of classical disorder and quantum fluctuations is further complicated by the computational challenges of performing even semiclassical calculations for disordered systems, where translational symmetry is broken, and traditional momentum space ({\bf Q}-space) approaches are not sufficient. This narrows the pool of theoretical techniques available to describe the dynamics. One approach is to employ spin wave theory (SWT), in the limit of a clean crystalographic cell, and treat the effect of disorder in perturbation theory through the self-energy of the magnon quasiparticles~\cite{Chernyshev02:65,Brenig91:43,Buyers72:5,Buyers73:6}. This is valid, for example, in the limit of small chemical disorder, but breaks down as one approaches the classical percolation threshold. An alternative, valid at larger defect concentrations but with the disadvantage of abandoning a quasiparticle description, is to perform classical simulations on finite systems in real space and finite temperature, using methods such as Monte Carlo (MC, for ${\mathcal{S}(\bf Q)}$) or Landau-Lifshitz Dynamics (LLD, for ${\mathcal{S}(\bf Q,\omega)}$). Provided the system is large enough, even with periodic boundary conditions, these classical calculations capture the effects of broken translational symmetry at large defect concentrations. However, these methods have several practical disadvantages: they are stochastic and require the acquisition of many samples to reduce statistical noise; they are limited in momentum space resolution by the lattice size $L$ along a given direction, $|\delta {\bf Q}| = 1/L$, making the simulation of high-dimensional systems challenging; and they cannot easily distinguish between effects from quenched and dynamical disorder, since the Fourier transform of the pair correlations and dynamical trajectories will contain contributions from both structural disorder and thermal fluctuations.

An alternative approach is to perform linear spin wave theory (LSWT) on large systems to accommodate particular disorder realisations. With an appropriate renormalization of the intensities, LSWT matches the LLD result in the $T\to 0$ limit. In contrast to classical dynamics, LSWT allows for control of the momentum resolution since spin-wave theory can be used to calculate observables at an arbitrary wavevector, benefits from the non-stochastic nature of spin-wave theory, and allows for a microscopic quasiparticle/operator-based understanding. However, the computational cost of spin wave theory is $\sim \mathcal{O}(N^3)$ is the number of sites $N$ in the system. This makes such calculations highly impractical for systems larger than $\approx$ 100 lattice sites. Recently, an implementation of spin-wave theory using the kernel polynomial method (KPM) was presented~\cite{Lane24:17}, which reduces the computational complexity to $\sim \mathcal{O}(N)$ in the system size, at the sole cost of losing the quasiparticle description. Similarly, real-space approaches such as Monte-Carlo and Landau-Lifshitz Dynamics can become highly impractical when large systems sizes are required, for instance, to achieve high momentum space resolution for clean but dynamically/thermally disordered systems. In that case, the Self-Consistent Gaussian Approximation (SCGA) has been used to accelerate the calculation of ${\mathcal{S}(\bf Q)}$ at the cost of neglecting anharmonic thermal effects. SCGA (and the equivalent Onsager Reaction Field Theory)~\cite{Conlon:thesis,Conlon10:81,Isakov04:93,Sen12:86,Gao17:13,Bai19:122,Gao22:128,Gao22:129,Brout67:3,Paddison20:125,Paddison23:35} assumes Gaussian fluctuations of the spin components and relaxes the spin normalisation condition, enforcing the spin length constraint only on average. This method is successful at modelling the spin correlations above $T_N$ in the paramagnetic phase.

In short, neutron scattering experiments are commonly used to search for quantum effects in magnetic systems. These experiments are necessarily conducted at finite temperatures and often involve systems that exhibit finite quenched disorder. This paper aims to demonstrate how semiclassical approaches can be employed to understand the spin dynamics of these systems while keeping computational costs modest. The ultimate goal is to enable the extraction of high-fidelity models of exchange interactions and disorder, and identify genuine quantum effects through the breakdown of semiclassical treatments. To illustrate recent progress, we apply various semiclassical methods, including the Kernel Polynomial Method with Spin-Wave Theory (KPM-SWT), Linear Spin-Wave Theory with direct matrix diagonalisation (LSWT), Landau-Lifshitz Dynamics (LLD), and the Self-Consistent Gaussian Approximation (SCGA), to several well-established systems dominated by quenched and dynamical disorder. The presented computational methods are available with a unified interface in the open-source Julia code based {\scshape Sunny.jl} ~\cite{Sunny:software}.

The paper is organised as follows. In Section~\ref{Section:QuenchedDisorder}, we explore several paradigmatic examples of quenched disorder (the dilute and random bond antiferromagnets). We confirm previous results derived analytically in the weak disorder limit, and extend these predictions to systems at, and above the classical percolation threshold. In Section~\ref{Section:DynamicalDisorder}, we explore clean systems (the pyrochlore antiferromagnet, pyrochlore slab and frustrated $J_1$-$J_2$ honeycomb) for which the Hamiltonian possesses translational symmetry, but the low energy states are disordered due to extensive, or large subextensive, ground state degeneracy or metastable local energy minima.

\section{Methods}
\label{Section:Methods}

Throughout this paper, unless otherwise noted, we calculate the dynamical spin structure factor (DSSF, $\mathcal{S}(\bf Q,\omega)$) of various models using the kernel polynomial method applied to linear spin wave theory (KPM-SWT) on large supercells. To connect with experimental observations, we project the spin-space components of the DSSF perpendicular to the momentum transfer, as is measured in unpolarized neutron scattering experiments. For the dilute systems in Sect.~\ref{Section:DiluteHeisenberg} and \ref{Section:RandomCoupling}, where the Hamiltonian possesses spin rotational symmetry, and we track the scattering intensities as a function of a disorder parameter, we instead plot the trace of the DSSF for consistency, since the spontaneous breaking of spin rotational symmetry is along an arbitrary direction. In real materials, this degeneracy is typically broken either by small terms in the Hamiltonian or higher-order contributions to the magnon self-energy, which are neglected in this analysis. 

Care must be taken when selecting the supercell size when performing supercell calculations. The imposition of periodic boundary conditions is needed to eliminate the resonance modes associated with the large number of dangling bonds at the system edges. However, this introduces a periodicity on the scale of the supercell size, leading to anomalies on the scale of $\sim 1/L$. To limit these artefacts, we will choose large values of $L$. For instance, for localisation to be visible, the system must be larger than the localisation length, $\ell$. Additionally, the system should be large enough to contain representative realisations of local environments. This requirement can often be fulfilled by summing calculations over independent realisations of the disordered configurations. 

In many cases, we compare our results to the relevant Linear Spin Wave Theory (LSWT), Self-Consistent Gaussian Approximation (SCGA), and Landau-Lifshitz Dynamics (LLD) calculations. All the results presented herein were obtained using the {\scshape Sunny.jl} codebase~\cite{Sunny:software} on an Intel Core$^{\mathrm{tm}}$ i7-1185G7 CPU, with the large supercell LSWT and KPM calculations parallelised on eight threads. LSWT and KPM calculations were performed using a Lorentzian kernel with a full width at half maximum of $2\Gamma$ with $\Gamma = 0.1JS$, except for the calculations in Sect.~\ref{Section:SCGO} for which $\Gamma = 0.25$~meV.

\section{Quenched Disorder}
\label{Section:QuenchedDisorder}
\subsection{Dilute Heisenberg Model on the Square Lattice}
\label{Section:DiluteHeisenberg}
The first model we consider is the dilute Heisenberg model on the square lattice,

\begin{equation}
\hat{\mathcal{H}}=\frac{1}{2}\sum_{ij}p_ip_jJ_{ij}\mathbf{\hat{S}}_i\cdot \mathbf{\hat{S}}_j \:,
\label{eq:diluteham}
\end{equation}
where the factors $p_i$ are unity unless $i$ is a vacancy site, in which case $p_i=0$. We consider a finite concentration of defects, $x$. At the classical percolation threshold, $x_p \approx 0.41$, there is no longer a connected network of spins, and long range magnetic order is destroyed~\cite{Sandvik02:66}. This model has attracted considerable interest due to its physical relevance to the cuprate superconductors, in which Cu$^{2+}$ ions form a square lattice of $S=1/2$ moments. Through substitutional doping, it has been suggested that La$_2$Cu$_{1-x}$Zn$_x$O$_4$ can be driven to a quantum critical point (QCP) prior to the classical percolation threshold~\cite{Chen00:61} as the effective dimensionality is reduced by the presence of non-magnetic vacancies which introduce frustration. However, careful experimental studies~\cite{Vajk03:126} and Quantum Monte Carlo~\cite{Sandvik01:86,Sandvik02:66} results have contradicted suggestions of a QCP below the classical percolation threshold. 

Close to the percolation threshold, quantum fluctuations, due to the formation of low dimensional clusters, are expected to affect the low energy dynamics considerably, altering the critical exponents and giving rise to different types of excitations including an Anderson quantum rotor mode and multimagnon continuum~\cite{Vojta05:95,Wang06:97,Wang10:21,Dao24:preprint}. These phenomena are absent from a semiclassical treatment and must be treated using techniques such as exact diagonalisation of a distribution of clusters.

The dynamics of this model (for $S=1/2$) have been explored extensively using the $T$-matrix formalism in the limit of dilute vacancies, $x \ll 1$~\cite{Brenig91:43,Chernyshev01:87,Chernyshev02:65}. We summarise the main findings of this previous work~\cite{Chernyshev02:65} below:
\begin{enumerate}
    \item A hydrodynamic description of excitations in terms of spin waves breaks down at wavevectors greater than a localisation energy scale $\ell/a\sim e^{\pi/4x}$ (expressed in dimensionless units of energy $E/zJS$).
    \item A low energy localisation peak is found at $\omega_0 =c_0/\ell$, where $c_0$ is the bare spin wave velocity.
    \item An abnormally damped quasiparticle peak exists between the spin wave branch of the clean system and the localisation peak.
    \item A flat background of states exists between the quasiparticle and localisation peaks.
    \item The quasiparticle and localisation peaks merge below $k \sim \ell/a$.
    \item A defect resonance mode is seen at $\omega \sim J$. 
\end{enumerate}
\begin{figure}
    \centering
    \includegraphics[width=1.\linewidth]{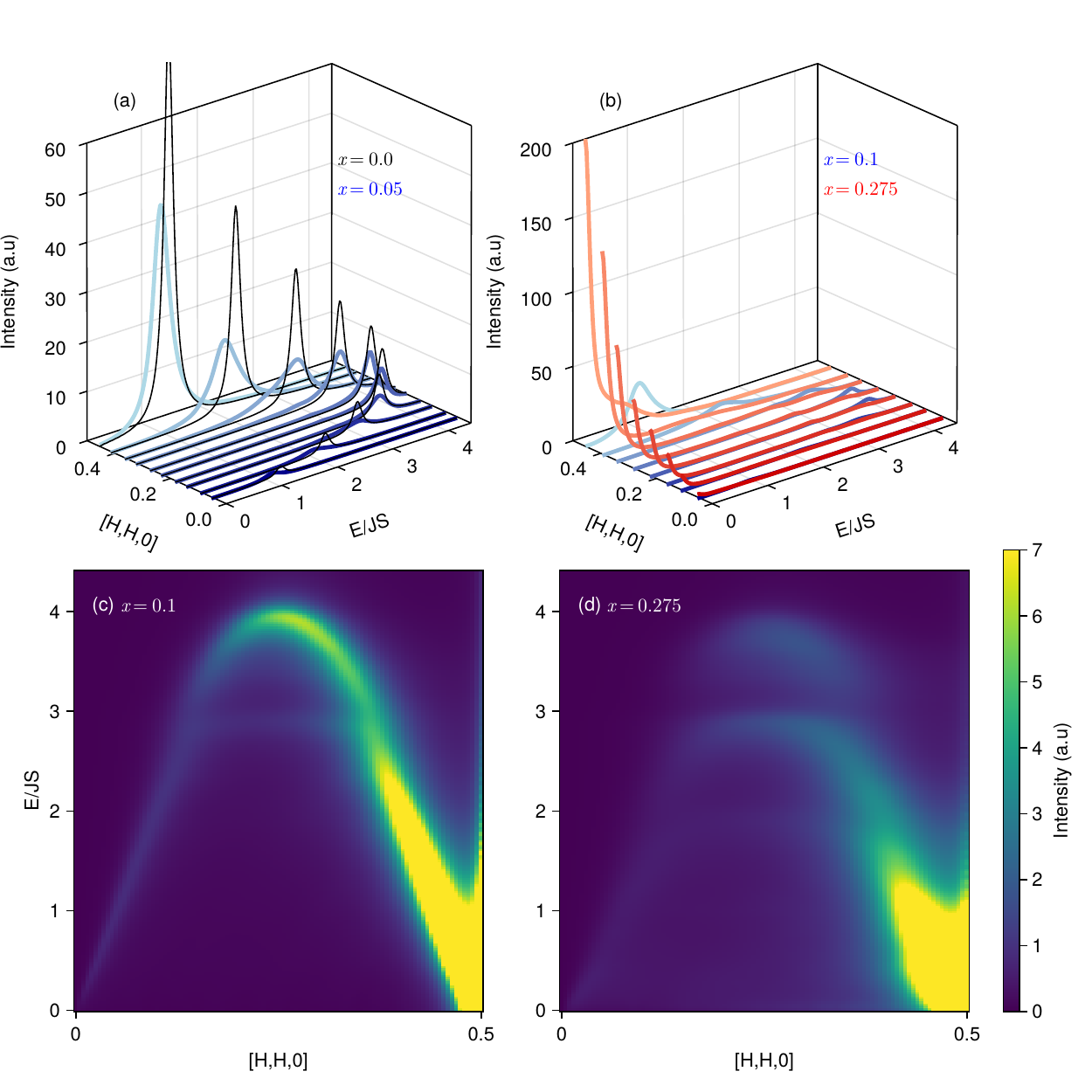}
    \caption{(a) One-dimensional constant-{\bf Q} cuts through the DSSF of the Hamiltonian of Eq.~\ref{eq:diluteham} in the weak dilution limit ($x=0.05$). The black line shows the magnon dispersion curve for the clean system. (b) One dimensional cuts for $x=0.1$ (blue) and $x=0.275$ (red) calculated using LSWT for an $80 \times 80$ supercell. A strong localisation peak is observed for $(x=0.275)$. Spectrum across for (c) $x=0.1$ and (d) $x=0.275$ showing clear evidence of a resonance peak.}
    \label{fig:Localisation}
\end{figure}
 Further details and discussion of these features can be found in Refs.~\cite{Chernyshev02:65,Dao24:preprint}. We now benchmark KPM-SWT to these analytical predictions by calculating the DSSF for large system sizes. We note that real-space spin wave theory calculations were already performed on realisations of dilute disorder in~Ref.\cite{Mucciolo04:69} in a moderately sized supercell $(32 \times 32)$. There, the largest connected cluster was identified for each realisation and spin wave theory was performed in real space on this cluster. As a result, the momentum resolution was coarse, limited by the supercell size. Moreover, the contribution from a distribution of clusters was not captured due to their systematic elimination from the problem, and the reduced supercell size precluded the observation of low-energy localisation modes.

Our calculations build on those of Ref.~\cite{Mucciolo04:69} by increasing the system size significantly to a $100 \times 100$ supercell and summing over all clusters within the supercell. This allows for a full accounting of the effect of cluster size distribution and provides the momentum resolution required to study fine details of both the low and high-energy parts of the spectrum. As discussed in Sect.~\ref{Section:Methods}, due to the spin rotational symmetry of the parent model, we plot the trace of the structure factor. The magnetic ground state was found by selecting a spatial distribution of defect sites, performing simulated annealing from high temperature to $T \rightarrow 0$~K, followed by energy minimisation using the conjugate gradient method. The DSSF is then obtained after averaging over three independent defect realisations. 

The simplest consequence of site dilution is the renormalisation of the spin wave energies. Such effects can be well accounted for in the dilute vacancy limit, $x \ll \ell$. Fig.~\ref{fig:Localisation}a shows a series of one-dimensional cuts through the spin wave spectrum, calculated using KPM-SWT for weak site dilution $x=0.05$. Plotted in black is the spectrum for the clean system. The quasiparticle peak is down-shifted in energy with a broadened linewidth. The spectrum is quantitatively similar to the clean system. We comment that instrumental resolution, along with contributions at higher order in $1/S$ make the experimental identification of renormalisation and damping due to quenched disorder difficult in the dilute vacancy limit. However, the renormalisation of the dispersion is not uniform throughout the Brillouin zone, with the quasiparticle peaks close to the zone centre experiencing more dramatic shifts than near the band maximum. The momentum dependence of the disorder-induced renormalisation was demonstrated analytically in Ref.~\cite{Chernyshev02:65} and presents challenges for fitting inelastic neutron scattering spectra to clean models as the effect cannot be simply captured by a global renormalisation of exchange parameters. We note that the scattering away from the $(\pi,\pi)$-point ($\mathbf{Q}=(0.5,0.5,0)$) is suppressed by the antiferromagnetic structure factor, making the identification of the features close to the crystallographic zone centre challenging.      

We now move to the question of magnon localisation. Careful analytical calculations have revealed the existence of low energy localised magnon modes on the length scale $\ell/a \sim e^{\pi/4x}$~\cite{Chernyshev02:65}. To observe significant magnon localisation, the system must be larger than the localisation length. However, anomalous localisation may exist where defects have clustered at a sufficient density by random. Such an effect can be mitigated by choosing a large system size and summing over multiple disorder realisations. The energy scale of this localisation peak is predicted to be $\sim e^{-\pi/4x}$, which is $\approx0.0004JSz$ for $x=0.1$. In general, KPM-SWT is not well-suited to quantitative studies of the spectral intensity at energies far below the bandwidth $\hbar \omega \ll \gamma$ since artefacts from destructive interference with negative energy eigenvalues can be significant there. Such effects can suppress the intensity of low-energy features. 

We therefore begin with standard LSWT for a limited number of wavevectors (on a smaller $80 \times 80$ supercell owing to the significant computational and memory cost associated with performing LSWT on large systems) before using KPM-SWT to map out the spectrum across the full Brillouin zone. To search for magnon localisation, we chose a doping concentration of $x= 0.275$. The analytical localisation length at this doping concentration is predicted to be $\sim 17$~\cite{Chernyshev02:65}, which comfortably fits within a $80 \times 80$ supercell. The analytical calculations of Ref. ~\cite{Chernyshev02:65} predict the localisation energy scale to be $\sim 0.06JSz$ and hence this should take the form of a quasielastic feature when convolved with a Lorentzian of finite width.

In Fig.~\ref{fig:Localisation}b, a series of one-dimensional constant-$\mathbf{Q}$ cuts through the spectrum are plotted, as calculated using LSWT averaged over three realisations of disorder (and $80 \times 80$ supercell). For $x=0.275$, a clear peak is seen at low energy associated with magnon localisation. Such dramatic features are absent at $x=0.1$ where the localisation length is predicted to exceed the system size. We now turn to the full spectrum across the entire Brillouin zone using KPM-SWT. In Fig.~\ref{fig:Localisation} (c,d) the spectrum is plotted across the Brillouin zone from the $\Gamma$ point to the zone corner. At both $x=0.1$ (c) and $x=0.275$ (d), a clear resonance peak is seen at around $\omega/J \approx 3$. At $x=0.275$ a continuum extends up to the band maximum, as predicted in Ref.~\cite{Chernyshev02:65}.
\begin{figure}
    \centering
\includegraphics[width=0.75\linewidth]{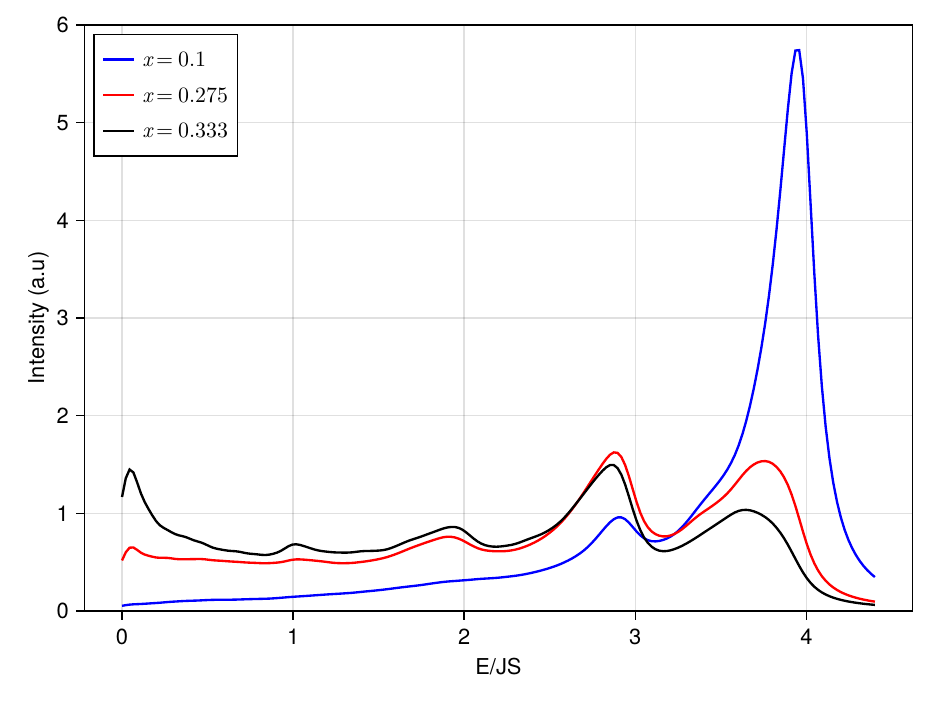}
    \caption{Cut at $\mathbf{Q}=[1/4,1/4,0]$ for varied defect concentrations. Calculation performed on a $100 \times 100$ supercell using KPM. The spectrum was averaged over three realisations}
    \label{fig:resonances}
\end{figure}

The resonance peak, so-called since it can be described by defect scattering contributions to the $T$-matrix, admits a qualitative interpretation as arising due to the reduction of the molecular mean field on sites adjacent to a defect. As such, away from the dilute limit, where vacancies are no longer spatially isolated, multi-defect resonances may occur. These occur in multiples of the vacancy energy $J$ and, in the molecular mean-field interpretation, originate from a finite density of spins with only one, two or three attached bonds. These excitations have been previously discussed in the context of the dynamics of fractal structures~\cite{Mucciolo04:69} -- localised excitations on percolating clusters which exist at higher energies than the propagating magnon mode~\cite{Orbach82:43,Orbach86:231,Nakayama94:66,Uemura86:57,Uemura87:36}. The origin of these additional modes is the reduced coordination and the degree to which they are localised depends strongly on the mode energy and vacancy concentration, with additional modes appearing well below percolation. Whilst KPM-SWT does not offer access to the eigenvectors of the dynamical matrix, it is possible to deduce the nature of the excitations by the degree to which they disperse and connect to fluctuations in the long wavelength limit. 

For defect concentrations $\gtrsim \pi/4\mathrm{ln}(L)\approx 0.2$, the localisation length is sufficiently small for magnon localisation to fit within a box of length $L=100$ along any direction. It is therefore possible to distinguish between those excitations localised on the scale of a cluster, which we term magnons, and those localised on shorter length scales. Fig.~\ref{fig:resonances} shows a cut through $\mathbf{Q}=[1/4,1/4,0]$ for different defect concentrations. A flat background of states is visible up to the quasiparticle peak for all concentrations, which intensifies on doping with vacancies. The primary (single-defect) resonance $\sim 3J$ is visible for $x=0.1$ with two and three defect resonances at $x=0.275$ and $x=0.333$. 

\begin{figure}
    \centering
    \includegraphics[width=1\linewidth]{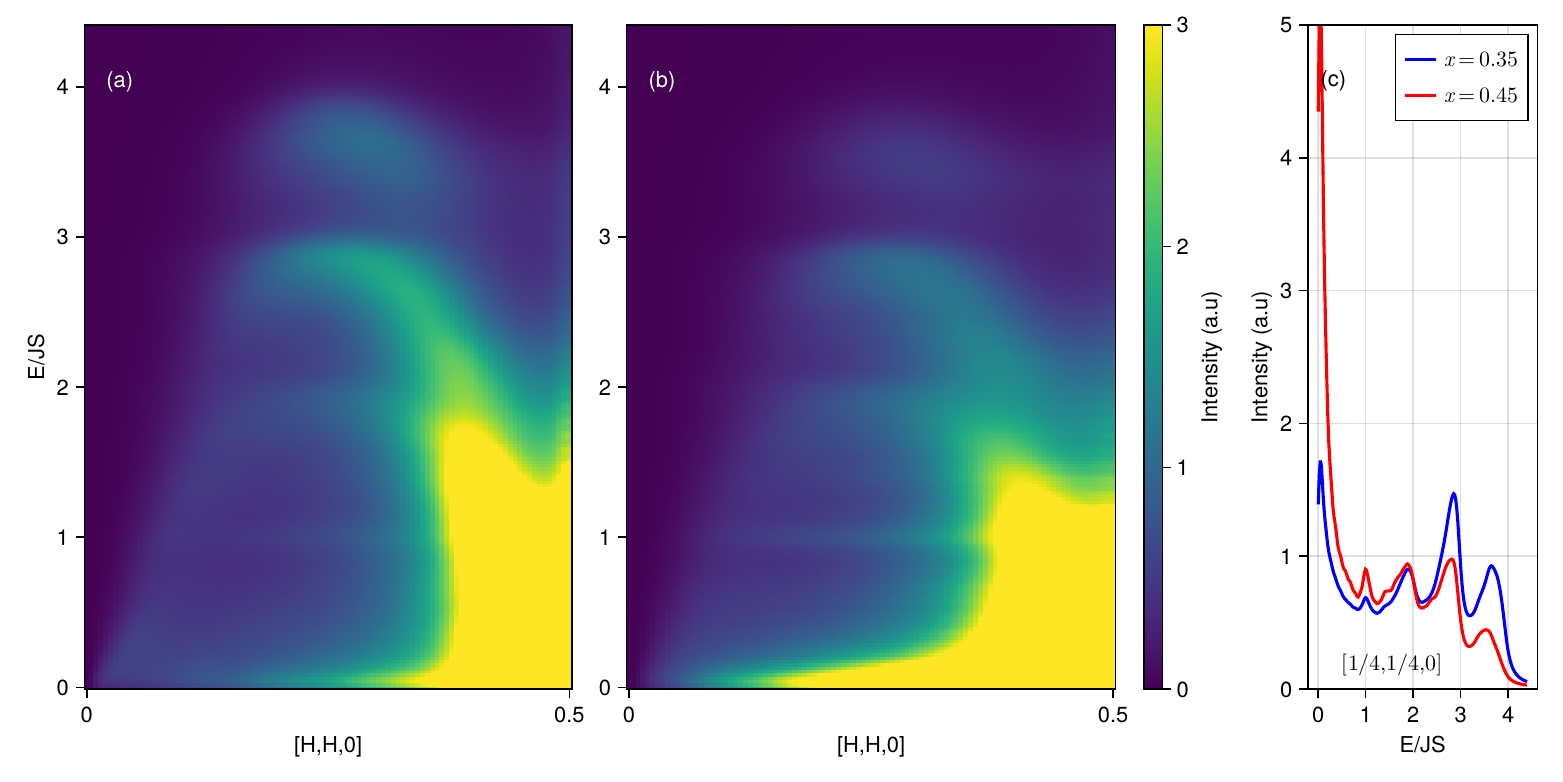}
    \caption{Spectrum of the dilute Heisenberg model on the square lattice at (a) $x=0.35$ and (b) $x=0.45$. (c) One dimensional cuts at $\mathbf{Q}=[1/4,1/4,0]$ demonstrating the persistence of resonance peaks above $x_p$.}
    \label{fig:percolation}
\end{figure}

Having demonstrated the utility of KPM-SWT for capturing effects due to multi-defect interactions, we extend the calculations to the large defect limit. Fig.~\ref{fig:percolation} shows the spectrum for $x=0.35$ and $x=0.45$, below and above the percolation threshold $x_{p} = 0.41$. The spectra look qualitatively similar, with intensity redistributed from the quasiparticle peak to a low-energy quasielastic feature. The shape of the spectrum close to ${\bf Q}=[0.5,0.5,0]$ also differs above and below $x_p$ with a mound of intensity that spreads to finite wavevector [Fig~\ref{fig:percolation}b], replacing a column of intensity which results from the ill-defined spin wave velocity [Fig~\ref{fig:percolation}a]. Constant $\mathbf{Q}$ cuts [Fig~\ref{fig:percolation}c] demonstrate the persistence of resonance modes above $x_p$ corresponding to defect resonance modes on finite-sized decoupled clusters. Above percolation, the low energy localisation mode appears to gain significant weight, with intensity redistributed from the peaks at $\omega = 3J$ and $\omega = 4J$ to lower energy.   
Enabled by the fine momentum resolution of KPM-SWT, we further comment that the low energy dispersive component of the spin wave branch smoothly connects to the single defect resonance for larger values of $x$. This is visible to some extent in Fig.~\ref{fig:Localisation}d but is most evident in Fig.~\ref{fig:percolation}a. This is indicative that for the largest connected cluster, each site is, on average, adjacent to a vacancy. The average molecular mean field, therefore, becomes $3J$, and around $x=1/4$, the resonance is not localised but magnon-like. This crossover from local resonance to magnon can be expected to occur at critical concentrations on other lattices as dictated by the coordination number, for example, at $x=1/3$ on the triangular or honeycomb lattices.

To conclude this section, we remark on several key findings. Our calculations agree with the analytical calculations of Ref.~\cite{Chernyshev02:65}, demonstrating the renormalisation of the spin-wave dispersion, along with the appearance of a localisation peak, flat continuum and a resonance mode. These calculations extend the previous analytical results beyond the dilute limit, showing the emergence of multi-defect resonance structures at large doping concentrations as observed in single largest connected cluster calculations as presented in Ref.~\cite{Mucciolo04:69}. Enabled by the fine momentum resolution provided by our large supercell calculations, we identified a crossover from localised resonances to extended magnons at $x=1/4$. The results of this section demonstrate the necessity for comparison of experimental data with semiclassical calculations for disordered systems. In particular, several features of our semiclassical calculations: the emergence of a broad continuum, quantised peaks, anomalous energy damping and a renormalised spin wave velocity are effects that could experimentally be mistaken for quantum mechanical phenomena such as fractionalised quasiparticles, bound-state formation, and nonlinear spin-wave corrections, respectively. Our results demonstrate the feasibility of calculations, with sufficient resolution to compare with experiments using KPM-SWT.

\subsection{Random Coupling}
\label{Section:RandomCoupling}
We now move our attention to a different form of quenched disorder, where each site is occupied by the same species of magnetic ion, but the interaction strengths follow some statistical distribution. This family of models has generated interest both from a materials-driven and a fundamental perspective~\cite{Lin03:68,Laflorencie06:73,Kruger02:65,Sandvik95:74}. Such a situation can occur due to nonmagnetic disorder of the surrounding crystallographic environment, for example, the presence of oxygen vacancies or site-mixing. We again will consider the square lattice Heisenberg model with nearest neighbour interactions 
\begin{equation}
    \mathcal{H} = J\sum_{\langle i,j\rangle} p_{ij}\mathbf{\hat{S}}_i \cdot \mathbf{\hat{S}}_j
\end{equation}
\noindent where $p_{ij}$ is randomly distributed according to some distribution. Perhaps the most experimentally relevant are normal~\cite{Edwards75:5,Sherrington75:35,Hikaru03:36} and bimodal~\cite{Toulouse77:2,Aharony80:13,Tang15:91} distributions. The former could be driven by subtle variations in orbital overlap due to structural disorder and the latter due to non-stoichiometry or oxygen vacancies, leading to the absence of the contribution from a particular exchange path along some fraction of the bonds in a given system. One bimodal distribution that has elicited attention is the random sign antiferromagnetic Heisenberg model~\cite{Toulouse77:2,Aharony80:13}. The presence of a finite number of ferromagnetic bonds introduces frustration, and it has been suggested that this can lead to the formation of a spin glass phase on some lattices~\cite{Edwards75:5,Binder86:58,Nonomura95:64,Oitmaa01:87}. Whilst at first glance this model seems contrived, such a situation may occur in doped systems, for example doped La$_2$CuO$_4$ where the presence of holes creates ferrromagnetic bonds in the parent antiferromagnetic system~\cite{Aharony88:60}. Another special case is -the bond dilution, where some fraction of the bonds are set to zero. This model has been studied through Monte Carlo in Ref.~\cite{Sandvik02:66} and shows the loss of order at the classical bond percolation threshold like the site dilution model studied in Sect.~\ref{Section:DiluteHeisenberg}. 
\begin{figure}[h]
    \centering
    \includegraphics[width=1\linewidth]{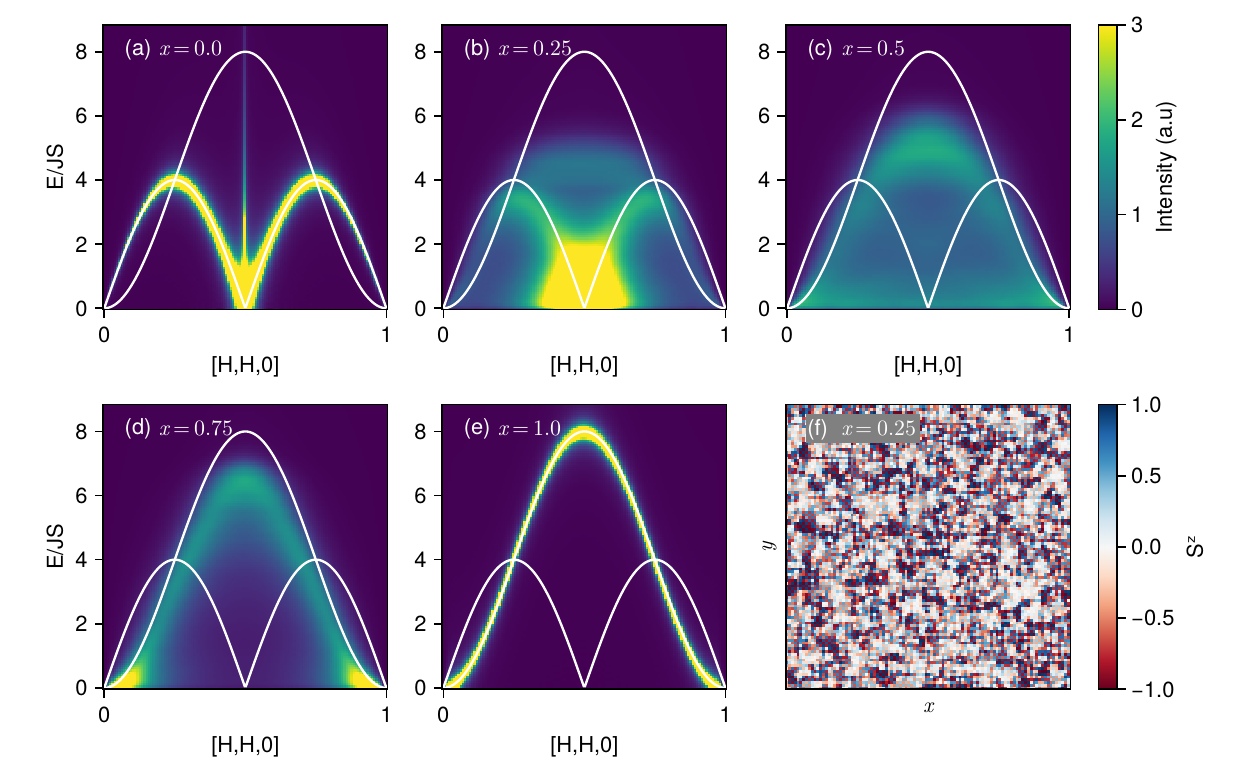}
    \caption{(a-e) DDSF for the random sign Heisenberg model as a function of doping, $x$. Overlaid are the dispersion curves for the clean ferromagnet and antiferromagnet. (f) Real-space spin configuration for $x=0.25$.}
    \label{fig:randomsign}
\end{figure}
To examine the behaviour of the classical model as a function of $x$ for different distributions, we perform KPM-SWT on a large supercell of size $100 \times 100$ with a finite concentration of ferromagnetic bonds, $x$. The ground state was determined by simulated annealing followed by gradient descent for three independent realisations of bond disorder. The DSSF was calculated for each realisation and summed to achieve a representative calculation of disorder in the thermodynamic limit. We emphasise here that the goal is not to find the global magnetic ground state, which is a formidable task for spin glasses and for some models is NP-hard~\cite{Barahona82:15}. Instead, we seek a low energy local minima around which we can perform linear spin wave theory. When performed in a large supercell and averaged over multiple realisations, this procedure is a good approximation for experiments performed at finite but low temperatures. 

\subsubsection{Random sign}
In the case of the quantum random sign Heisenberg model on the square lattice, with the distribution
\begin{equation}
    P(p_{ij};x)=x\delta (p_{ij}+1) + (1-x)\delta (p_{ij}-1),
\end{equation}
\noindent previous studies of the quantum problem have estimated critical density of ferromagnetic bonds to be $x_c = 0.11$ to destroy N{\'e}el order, after which there is some evidence of spin glass behaviour via exact diagonalisation~\cite{Nonomura95:64,Oitmaa01:87}. Sign-problem-free quantum Monte Carlo calculations suggest that small percentages of ferromagnetic bonds suppresses the long-range antiferromagnetic order~\cite{Sandvik94:50}. Here, we address the dynamics of this model using semiclassical methods.

Fig.~\ref{fig:randomsign} shows the spectrum as a function of doping, $x$ from the clean antiferromagnet $(x=0)$ to the clean ferromagnetic $(x=1)$ limit. On the addition of ferromagnetic bonds, a continuum of scattering emerges both above and below the magnon dispersion. A large mound of intensity is observed close to the putative antiferromagnetic wavevector [Fig.~\ref{fig:randomsign}b]. A damped remanent magnon dispersion curve is still seen despite the loss of long range order. This is reflective of the presence of small regions of antiferromagnetic order [Fig.~\ref{fig:randomsign}f]. At the symmetric point, $x=0.5$, intensity is seen throughout the Brillouin zone, peaking at $\omega/JS = 6$.at the zone boundary. The intensity is remarkably constant throughout the mound, with a slight peak around the continuum maximum. In the ferromagnetically-rich model ($x=0.75$), a damped ferromagnetic magnon response is observed, but the excitations are still relatively coherent. 
\subsubsection{Gaussian distribution }
We now consider the case of Gaussian distributed exchange constants. We draw from a distribution centred on $J$ with standard deviation, $\sigma$. A $100 \times 100$ supercell was used and the ground state determined by simulated annealing followed by gradient descent. The calculation was summed over three realisations of this large supercell. 
\begin{figure}[h]
    \centering
\includegraphics[width=1.0\linewidth]{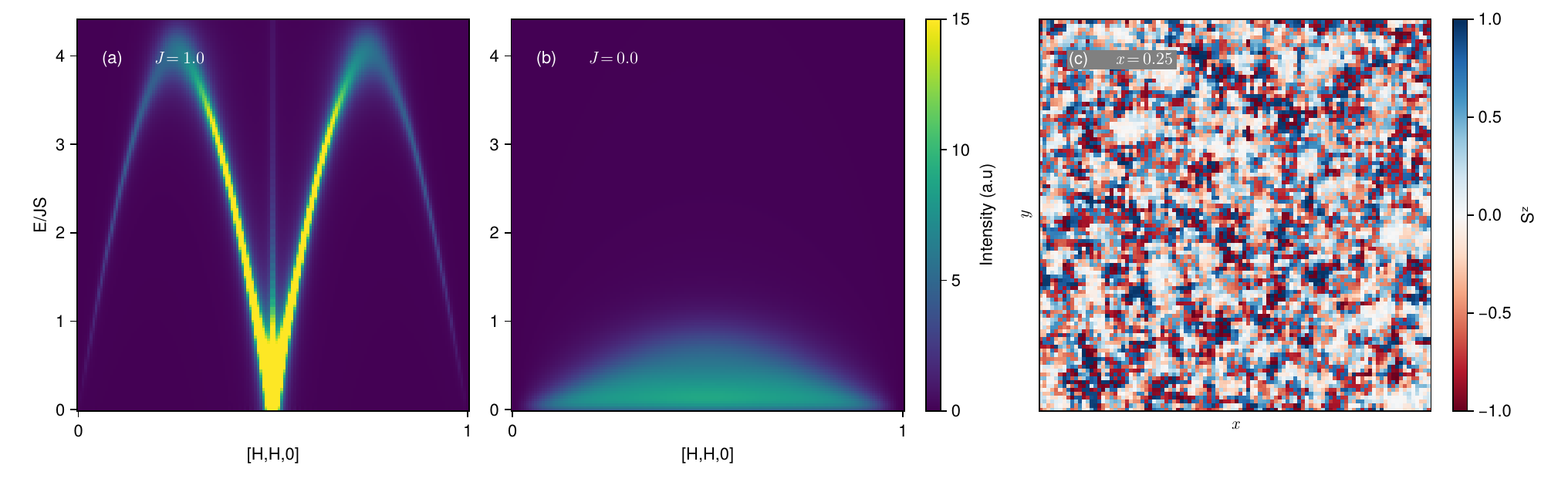}
    \caption{Spectra of the bond disordered model with a Gaussian distribution of exchange constants with mean (a) $J=1.0$ and (b) $J=0.0$ and standard deviation $\sigma=1/8$. (c) A low energy state of the the $J=0.0$ model which is disordered.}
    \label{fig:randomgaussian}
\end{figure}
Fig.~\ref{fig:randomgaussian} shows the DSSF plotted in the $[H,H,0]$ direction. With an antiferromagnetic coupling $J=1$, normally distributed with standard deviation $\sigma = 1/8$ [Fig.~\ref{fig:randomgaussian}a], the excitations have a well-defined dispersion. In the long-wavelength limit, the excitations are long-lived, with increased damping near the band maxima. There is no evidence of renormalisation of the dispersion and the band maxima coincide with that of the clean model. This is perhaps unsurprising since the average exchange constant is unchanged, $\langle J \rangle = 1$. Fig.~\ref{fig:randomgaussian}b shows the spectrum for the model with $J=0$ and $\sigma = 1/8$. The system exhibits no long-range order [Fig.~\ref{fig:randomgaussian}c], and the spectrum shows no coherent spin wave excitations. Instead a gapless mound of intensity is seen up to $\sim 1$ meV. The bandwidth is governed by the width of the exchange strength distribution, which is Gaussian with a full width at half maximum of $\sim 2.4\sigma$ leading to an approximate bandwidth of $\sim 9.6\sigma$.

\section{Dynamical Disorder}
\label{Section:DynamicalDisorder}
In section~\ref{Section:QuenchedDisorder}, we discussed the effects of quenched disorder on the low-temperature spin-wave dynamics of (initially unfrustrated) Heisenberg antiferromagnets. In this section, we start from frustrated models where the Hamiltonian is translationally invariant but the magnetic sector displays an extensive (or large subextensive) classical ground state degeneracy leading to the stability of a correlated paramagnetic state down to very low relative temperatures. These models have attracted attention to stabilize quantum spin liquids (QSLs)~\cite{Savary16:80} -- phases dominated by dynamical quantum disorder -- although order-by-disorder phenomena may often play a role. Broad continua of fractionalised quasiparticle in ${\mathcal{S}(\bf Q,\omega)}$ surviving in the limit of very low temperatures are the spectral signatures of QSLs, along with reciprocal space modulations of the ${\mathcal{S}(\bf Q)}$ in the absence of magnetic Bragg peaks. But correlated paramagnets can give rise to some of the identical signatures in ${\mathcal{S}(\bf Q)}$ and ${\mathcal{S}(\bf Q,\omega)}$. Here, the temperature can replace quantum fluctuations and allow a system to explore the quasi-degenerate manifold of low-energy states. We thus contrast the phenomenon of $thermal$ dynamical disorder with $quantum$ dynamical disorder. To demonstrate this effect, we perform KPM-SWT in large supercells of representative systems to examine their dynamics. In addition to KPM-SWT, we present LLD and SCGA calculations to explore features in the static structure factor, ${\mathcal{S}(\bf Q)}$.

\subsection{Classical Spin Liquid: Pyrochlore Antiferromagnet}
\label{Section:CSL}
One of the simplest systems for which the Hamiltonian has translational symmetry, but whose classical ground state displays no long-range order in the zero-temperature limit is the pyrochlore Heisenberg antiferromagnet,
\begin{equation}
    \mathcal{\hat{H}}= J\sum_{\langle i,j \rangle}\mathbf{\hat{S}}_i \cdot \mathbf{\hat{S}}_j. 
\end{equation}
\begin{figure}[h]
    \centering
    \includegraphics[width=1.0\linewidth]{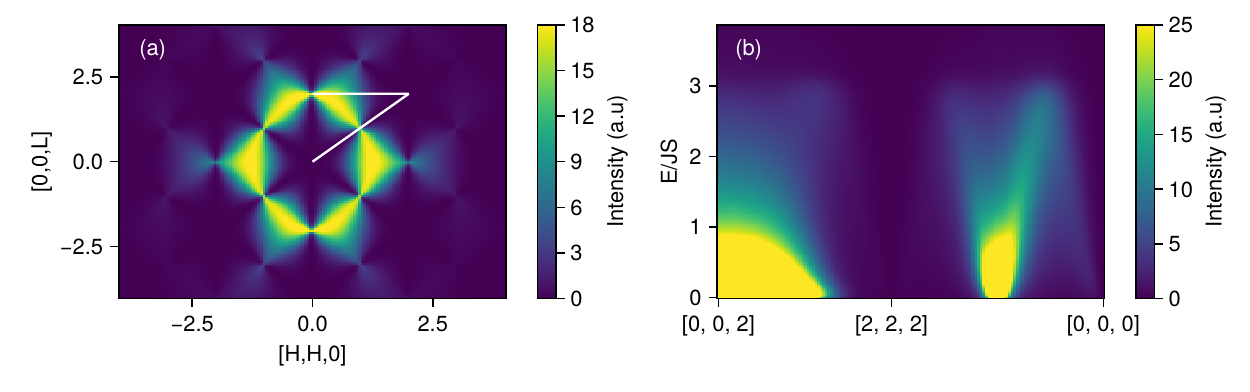}
    \caption{(a) Static structure factor for the antiferromagnetic pyrochlore model calculated using the Self Consistent Gaussian Approximation (SCGA). (b) Energy-resolved neutron scattering response along the white path shown in (a).}
    \label{fig:CSL}
\end{figure}
\noindent Owing to geometric frustration, the pyrochlore antiferromagnet has an extensive ground state degeneracy, with the classical energy minimised by any state where the total spin on each tetrahedron sums to zero
\begin{equation}
    \sum_{i \in t}\mathbf{S}_i = 0. 
    \label{eq:CSL_GS}
\end{equation}
\noindent It is possible to interpret the spins as three component flux variables whence, in the continuum limit, the classical energy minimisation condition becomes a divergence-free condition, $\nabla \cdot \mathbf{B}(\mathbf{r}) = 0$, in analogy to Gauss' law for magnetism. Owing to this mapping onto a description in terms of flux variables and the connection to electromagnetism, this is termed the Coulomb phase and emerges in a variety of geometrically frustrated models~\cite{Conlon10:81,Henley10:1,Conlon:thesis}. The extensive ground state degeneracy of this simple model gives rise to a classical spin liquid, lacking long-range magnetic order. However, the ground state is highly correlated. Signatures of these correlations are visible in the static structure factor. Fig.~\ref{fig:CSL}a shows the static structure factor $S(\mathbf{Q})$ for the nearest neighbor antiferromagnetic Heisenberg model on the pyrochlore lattice, calculated using SCGA. Sharp pinch points are seen at positions (for example at $\mathbf{Q}=[0,0,\pm2]$ and $\mathbf{Q}=[\pm2,\pm2,0]$) and a highly structured momentum dependence is observed in the scattering pattern. The pinch point, which is a feature of the divergence-free condition [Eqn.~\ref{eq:CSL_GS}], is asymmetric in momentum, with no intensity along the high symmetry nodal lines and high intensity perpendicular to this direction [Fig.~\ref{fig:CSL}a]. This asymmetry is representative of the nature of the correlations along these two directions, with the width of the pinch point scaling as the inverse correlation length at finite temperature~\cite{Conlon:thesis}. Along $[0,0,L]$ and $[H,H,H]$, the pinch point is extremely sharp at $T\to 0$ reflecting long-range correlations along the direction associated with the nodal line. This is a consequence of the divergence free condition which necessitates that the total spin on adjacent $(100)$ layers be vanishing which in turn leads to an antiferromagnetic ordering of the $(100)$ planes~\cite{Moessner98:58}. In contrast, the broad nature of the pinch point in the transverse direction indicates short range correlated behaviour along within these planes. This can be seen more clearly by comparing the inelastic spectrum along both directions. Along $[H,H,H]$ coherent spin-wave-like excitations can be seen, whereas along $[H,H,0]$ a broad continuum of excitations is seen, as expected for a short-range correlated liquid [Fig.~\ref{fig:CSL}].

\subsection{Spin Glass Transition in Classical Spin Liquid}
\label{Section:SCGO}
We now consider the interplay of this dynamical disorder with quenched disorder of the type discussed in Sect.~\ref{Section:QuenchedDisorder}. It has long been known that geometrically frustrated magnets, such as the pyrochlore antiferromagnet, are unstable to the formation of glassy states at small concentrations of defects~\cite{Binder86:58,Ramirez94:24}. Two physically realisable scenarios are (i) the presence of nonmagnetic impurities and (ii) bond disorder. The thermodynamics of the former scenario has been extensively studied in the literature~\cite{Syzranov22:13,Syzranov22:106,Sen11:106,LaForge13:110,Schiffer97:56,Wollny11:107} and finds strong experimental motivation among the off-stoichiometric A$_{2}$B$_2$O$_7$ pyrochlore compounds~\cite{Gaudet16:94,Andrade18:120,Shirai17:96}. The case of bond-disorder (ii) has also attracted both theoretical~\cite{Andreanov10:81,Garratt20:101,Shinaoka11:107} and experimental interest~\cite{Booth00:62,Ratcliff02:65,Sarkar17:96} and arises due to the disordering of the local crystalline electric field by oxygen or other anion vacancies. Perhaps the most extensively discussed spin-glass system is the dilute system SrCr$_{9p}$Ga$_{12-9p}$O$_{19}$ (SCGO). The pure $p=1$ structure comprises kagom{\'e} bilayers and isolated spin dimers layered alternately. Samples with $p=1$ are unstable and the composition $p=8/9$, with nonmagnetic Ga$^{3+}$ occupying a small fraction of the Cr$^{3+}$ sites, is favoured. There is a small substitutional bias in the site dilution leading to fewer nonmagnetic vacancies at the $2a$ Wyckoff site which lies on the vertex shared by both tetrahedra in the pyrochlore slab~\cite{Limot02:65}. The coupling between each of the subsystems is weak, with the dimers contributing a high energy singlet-triplet mode that can be neglected from the analysis of the low energy physics. 

A minimal model, inspired by SCGO, formed of slabs of corner-sharing tetrahedra (pyrochlore slab) can then be constructed. We neglect the distortion of the tetrahedra present in SCGO and consider the bonds of the tetrahedra to be equivalent, following Ref.~\cite{Limot02:65}.
\begin{figure}[h]
    \centering
    \includegraphics[width=1.0\linewidth]{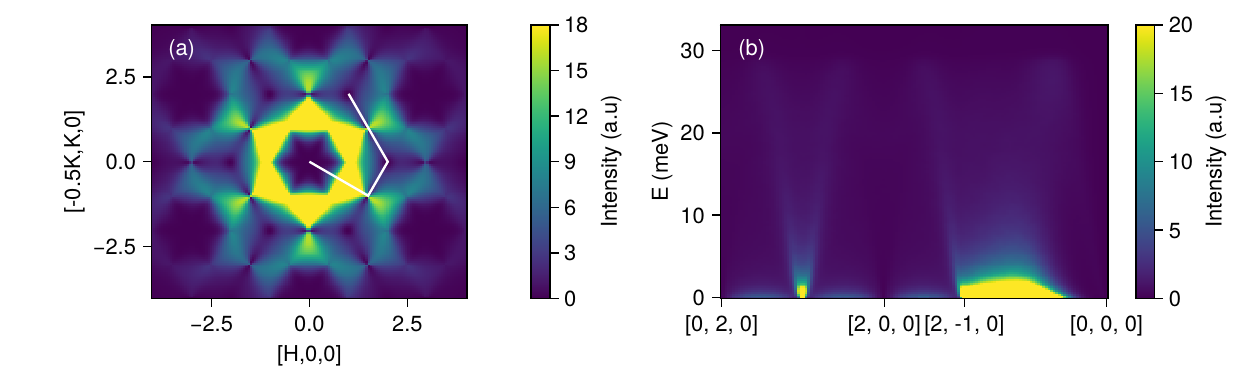}
    \caption{(a) Static structure factor for the SCGO-inspired model calculated using SCGA. (b) Cut through energy resolved neutron scattering response along the white path in (a), calculated using KPM-SWT.}
    \label{fig:SCGOclean}
\end{figure}
The ground state of the pyrochlore slab obeys a divergence free condition, leading to pinch points in the static structure factor [Fig.~\ref{fig:SCGOclean}a], which broaden as a function of temperature. This local constraint, as for the conventional pyrochlore lattice is indicative of U(1) gauge field. Fig.~\ref{fig:SCGOclean}b shows a cut along the path shown in Fig.~\ref{fig:SCGOclean}a. As seen in the case of the pyrochlore antiferromagnet, cuts along the nodal lines show sharp spin-wave like excitations whilst in the transverse direction, a broad mound of intensity is seen.

The presence of defect spins might well be viewed as excitations of the $U(1)$ gauge field since they lead to local violations of the local spin constraint. However, unlike the dynamical violations of the flux constraint, quenched defects are locked into place. Previous theoretical investigations~\cite{Sen11:106,Sen12:86} have shown that ``orphan spins" which live on a triangle with two spin vacancies carry fractional spin $S/2$. These ghost spins or quasispins contribute to the magnetic susceptibility and offer an explanation for the increased susceptibility of spin liquid systems as a function of defect concentration~\cite{Syzranov22:13}. 

We now consider the dynamics of SCGO in the spin glass phase. Owing to the loss of translational symmetry in the Hamiltonian, it is no longer practical to calculate the static structure factor using SCGA, which would require the diagonalisation of a matrix that scales with the system size. Instead, we use LLD to simulate the dynamics, coupled to a thermal bath at $T = 0.1$~K. We consider a fraction of nonmagnetic defects $(1-p)$ randomly distributed throughout the lattice, and neglect the preference for defects to sit on the kagom{\'e} bilayers for simplicity.

\begin{figure}
    \centering
    \includegraphics[width=1.\linewidth]{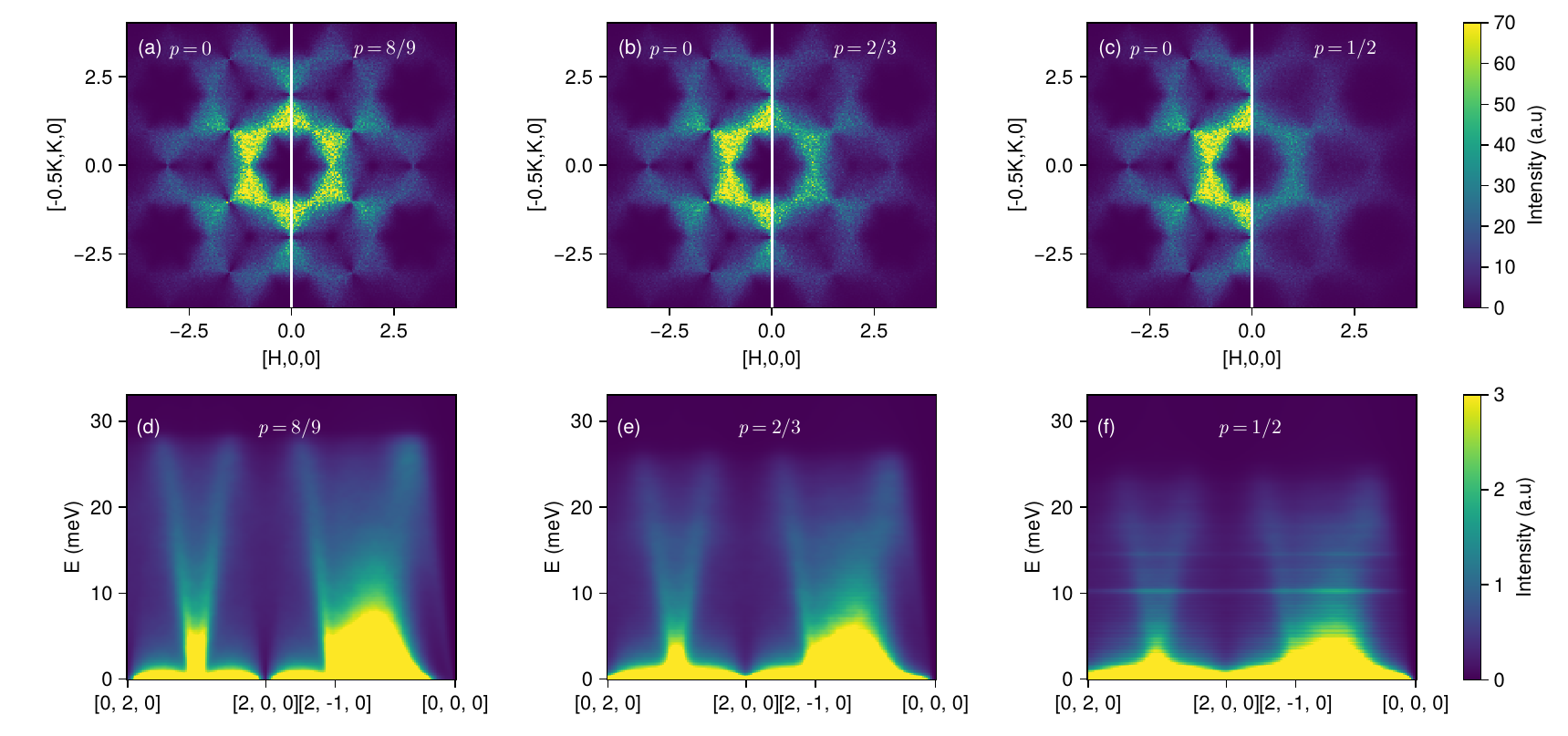}
    \caption{Static structure factor for the pyrochlore slab model calculated with LLD for nonmagnetic defects with (a) $p=8/9$, (b) $p=2/3$ and (c) $p=1/2$. Dynamical spin structure factor for pyrochlore slab model calculated with KPM-SWT with nonmagnetic defects for (d) $p=8/9$, (e) $p=2/3$ and (f) $p=1/2$. }
    \label{fig:SCGO-defect}
\end{figure}

Figs~\ref{fig:SCGO-defect}a-c show the static structure factor for $p=8/9$, $p=2/3$ and $p=1/2$ respectively, compared with the structure factor of the clean limit (left). The introduction of defects broadens the pinch point scattering as the presence of a vacancy causes a violation of the local spin constraint. Although significantly broadened, the momentum dependence of $S(\mathbf{Q})$ is remarkably persistent under doping  -- even up to $p=1/2$. 
The excitation spectrum is similarly robust to doping of nonmagnetic defects. As the fraction of defects grows, intensity is redistributed from high energies to quasi-elastic features [Fig.\ref{fig:SCGO-defect}d-f]. The spin-wave-like excitation branches are broadened as doping concentration increases. Remarkably, at very high defect concentrations, $p=1/2$, flat, localised resonance modes are observed at $\sim 10$ meV ($\approx JS$) and $\sim 15$ meV [Fig~\ref{fig:SCGO-defect}f], as observed in the case of the dilute square lattice Heisenberg antiferromagnet in Sect.~\ref{Section:DiluteHeisenberg}. 

In this section we have studied the pyrochlore slab model, inspired by the spin glass system SCGO. In the clean system, we observe sharp pinch points associated with an emergent $U(1)$ gauge field and excitations that show a broadened spin wave branch along the nodal line with diffuse energy broadened scattering in the transverse direction. The introduction of non-magnetic defects was observed to broaden the pinch points, consistent with the violation of the divergence-free condition, but the overall momentum dependence was remarkably unchanged. In the inelastic spectrum, intensity was redistributed to low energy quasielastic features and sharp, flat resonance modes were observed at large doping concentrations.

\subsection{Spiral Spin Liquid: $J_1$-$J_2$ Honeycomb }
In the previous sections, we presented two models of classical spin liquids for which the ground state has an extensive degeneracy. A related class of models exist with a \textit{subextensive} degeneracy. For these models, the energy is minimised along a contour or surface of wavevectors. In these systems, ground states are not connected by local spin flips~\cite{Yan22:4}, instead requiring global spin rotations to change the ordering wavevector $\boldsymbol{k}$, hence the low energy dynamics differ considerably between the two cases.

The route towards spiral spin liquid ground states is the construction of a model for which the classical energy is minimised by some contour or surface of magnetic ordering wavevectors. In Ref.~\cite{Niggemann20:32} a large family of models which host spiral spin liquid ground states on bipartite lattices in one and two dimensions were constructed based on the minimisation of the exchange matrix $J(\mathbf{q})$, in the spirit of the Luttinger-Tisza method~\cite{Luttinger46:70,Lyons60:5}. In particular, the models for which experimental realisations have been found include the honeycomb lattice~\cite{Gao22:128}, the diamond lattice~\cite{Gao17:13,Graham23:130} and the pyrochlore lattice~\cite{Gao22:129,Bai19:122}. 

In this section we consider the spiral spin liquid on the honeycomb lattice, inspired by the work of S. Gao \textit{et al} in Ref.~\cite{Gao22:128}. In Ref.~\cite{Gao22:128}, the authors presented neutron scattering measurements of the van der Waal's magnet FeCl$_3$ which has frustrated couplings $|J_1/J_2|\approx 0.36$. On the honeycomb lattice, for $|J_1/J_2| >1/6$, collinear N{\'e}el or ferromagnetic order is replaced by spiral contours~\cite{Okumura10:79,Mulder10:81}. The Hamiltonian for FeCl$_3$ is ultimately more complicated than a simple $J_1$-$J_2$ Heisenberg model, with interlayer and further neighbour interactions present which break the degeneracy of the spiral contour and select the commensurate wavevector $\boldsymbol{k}=(\frac{4}{15},\frac{1}{15},\frac{3}{2})$~\cite{Gao22:128}, however in this work we concern ourselves only with the dynamics about the spiral spin liquid ground state, and select $|J_1/J_2| =0.25$ with ferromagnetic $J_1$ and antiferromagnetic $J_2$.

The $J_1$-$J_2$ honeycomb lattice spiral spin liquid was explored analytically in Refs.~\cite{Okumura10:79} and \cite{Mulder10:81}. The classical energy is minimised for~\cite{Mulder10:81} 
\begin{equation}
    \mathrm{cos}(q_a)+ \mathrm{cos}(q_b)+\mathrm{cos}(q_a+q_b) = \frac{1}{2}\left[ \left(\frac{J_1}{2J_2}\right)^2-3\right].
\end{equation}
\noindent It has been suggested that for small $S$, the spiral spin liquid may be lost through quantum order-by-disorder~\cite{Mulder10:81}, and even in the absence of quantum fluctuations, the presence of small interaction terms such as long-range dipole-dipole interactions and further neighbour couplings likely select a state from the subextensive manifold of degenerate states in any real material as $T\to 0$. In numerical simulations, the presence of distinct liquid and glassy regimes have been demonstrated~\cite{Yan22:4}. The spin configuration locks to wavevectors that lie exactly on the degenerate spiral contour, with a density of topological defects at the intersection of domain walls.

\begin{figure}
    \centering
    \includegraphics[width=1.0\linewidth]{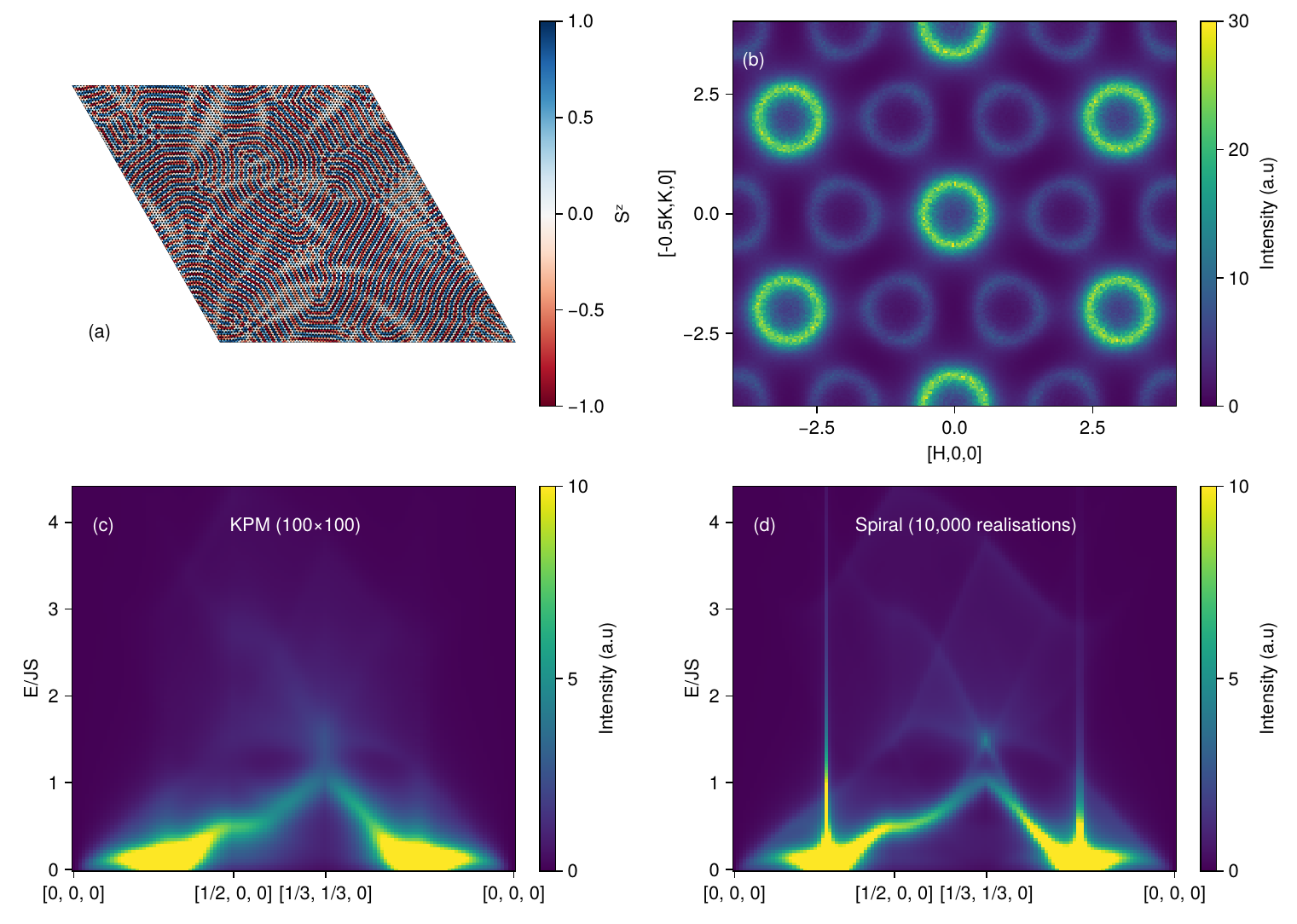}
    \caption{(a) Real space spin structure of one realisation of the SSL state (the colour of each hexagon represents the $S_z$ spin component). (b) Static structure factor for the state plotted in (a) at $k_{\rm B}T/J=0.1$. Dynamical spin structure factor along a high symmetry path calculated (c) on a large system size using KPM-SWT and (d) by averaging over the spectra for 10,000 realisation of single-$\boldsymbol{k}$ spiral order. }
    \label{fig:SSL}
\end{figure}

To explore this model we create three realisations of low energy configurations in a $100 \times 100$ supercell, cooled to $k_{\rm B}T/J=0.1$ through simulated annealing. It was then confirmed using LLD, that $S(\mathbf{Q})$ exhibited rings, indicating that the system is in a spiral spin liquid state. In order to perform linear spin wave theory, a local energy minimum was then found in close proximity to the state found through simulated annealing, by performing gradient descent. It was the verified that the ground states found after gradient descent [Fig.~\ref{fig:SSL}a] still produced closed contours in $S(\mathbf{Q})$ [Fig.~\ref{fig:SSL}b]. The real space spin configuration is clearly highly correlated but exhibits a large concentration of domain walls.  

Fig.~\ref{fig:SSL}c shows the dynamical structure factor through a high symmetry path through the Brillouin zone. High intensity scattering is seen at low energy as the path passes through the contour of degenerate wavevectors. A spin wave branch is clearly seen extending up to $\omega = JS$. Weak intensity is clearly visible up to energies $\omega = 4JS$, with a broad continuum-like character.

To explore the consequences of the presence of domain walls, we compare our large box KPM-SWT results [Fig.~\ref{fig:SSL}c] to an average over the spectrum of 10,000 realisations of a single $\boldsymbol{k}$ spiral magnetic structure where $\boldsymbol{k}$ is selected from the manifold of degenerate wavevectors. Each realisation was generated by randomly an initial guess of the ordering wavevector and a normal vector to the spin rotation plane, and minimising the classical interaction energy $J(\mathbf{q})$ using gradient descent methods. The spectrum was then calculated for each of these realisations using conventional rotating frame spin wave theory~\cite{Toth15:27}. The averaged single-$\boldsymbol{k}$ spiral spectrum [Fig.~\ref{fig:SSL}d] is qualitatively similar to that of the KPM-SWT calculation [Fig.~\ref{fig:SSL}d]. The effect of domain walls and topological defects is to broaden the spectrum at all energies. As a result, some of the sharp features such as the linear band crossing at the Dirac point are not visible in the KPM-SWT calculation but show up clearly in the spiral calculation. 

\section{Conclusion}
In this paper, we have analysed the static and dynamic spin correlations of several clean and disordered spin models using an array of semi-classical approaches, including LSWT, KPM-SWT, SCGA and LLD. Our study demonstrates that many features captured by neutron scattering experiments and associated at face value with quantum mechanical phenomena, such as the emergence of continuous excitations, quantised modes and quasiparticle damping, can also be present in semi-classical calculations, in which case they stem from classical phenomena. We have made the distinction between quenched disorder and dynamical (thermal) disorder, both of which can be modelled using semiclassical techniques on large system sizes as demonstrated in Sect.~\ref{Section:QuenchedDisorder} and Sect.~\ref{Section:DynamicalDisorder} respectively. Our work demonstrates the feasibility of calculating the dynamical spin structure factor for spin systems of underlying sizes of $\sim$ 100,000 sites, without the need for high-performance computing resources. Given the low cost of these calculations, our results highlight the importance of comparing experimental data with semiclassical models, including realistic models of quenched and thermal disorder, to exclude a classical origin in the search for quantum disordered phases of matter. 

\section{Acknowledgements}
We would like to thank A. L. Chernyshev for inspiring us to examine the dilute Heisenberg model and for useful conversations. We acknowledge useful discussions with C. D. Batista. H. Lane would like to acknowledge financial support from the Royal Commission for the Exhibition of 1851. The work at Georgia Tech (M. Mourigal) and LANL (K. Barros) was supported by U.S. Department of Energy, Office of Science, Basic Energy Sciences, Materials Sciences and Engineering Division under awards DE-SC-0018660 and DE-SC0022311, respectively. The research data supporting this publication can be accessed at http://dx.doi.org/\#\#\#~\cite{Data:data}.
\section*{References}
\bibliographystyle{iopart-num}
\bibliography{references.bib}
\end{document}